\documentclass[
  groupedaddress,
  amsmath,amssymb,
  prb,
  floatfix,
  superscriptaddress,
  twocolumn,
  aps
]{revtex4-2}

\usepackage{graphicx}
\usepackage{dcolumn}
\usepackage{bm}
\usepackage[mathlines]{lineno}
\usepackage{xcolor}
\usepackage{soul}
\usepackage{adjustbox}
\usepackage{nth}
\usepackage[T1]{fontenc}
\usepackage{nth}
\usepackage{amsmath}
\usepackage{amssymb}
\usepackage{hyperref}
\usepackage{cleveref}
\usepackage{xr}

\begin{document}

\title{Embedded Cluster Expansions for Exploring Diffusion in Compositionally Complex Materials}
\author{Damien K. J. Lee}
\affiliation{Laboratory of materials design and simulation (MADES), Institute of Materials, \'{E}cole Polytechnique F\'{e}d\'{e}rale de Lausanne}
\author{Anirudh Raju Natarajan}
\email{anirudh.natarajan@epfl.ch}
\affiliation{Laboratory of materials design and simulation (MADES), Institute of Materials, \'{E}cole Polytechnique F\'{e}d\'{e}rale de Lausanne}

\begin{abstract}
  Diffusion in compositionally complex materials is difficult to predict because the large number of distinct local chemical environments produces a wide distribution of migration barriers, and thermodynamic and kinetic effects are coupled. Cluster expansions of the migration barriers in these materials contain many coefficients and require large first-principles training sets. Here, we present an open-source implementation of the embedded local cluster expansion (eLCE) framework coupled to kinetic Monte Carlo simulations in the \texttt{pyeCE} package. Chemical embedding reduces the amount of training data that eLCE models require. Models trained only on binary and ternary configurations extrapolate to quaternary through senary alloys with errors of about 0.1\,eV, and models with test mean absolute errors as large as 190\,meV reproduce the qualitative trends in the diffusion quantities. We apply these models to bcc alloys of the group 5 and 6 elements V, Nb, Ta, Cr, Mo, and W. In NbMoTaW at 1000\,K, short-range order raises the migration barriers of Nb and Ta and lowers their tracer diffusion coefficients by factors of about 6 and 8 relative to the random alloy. We quantify sluggish diffusion with elemental enhancement factors relative to the pure constituents, which range from about $10^{-3}$ in NbMo to about 10 in CrW. Separating these factors into vacancy, kinetic, and correlation contributions shows that the vacancy concentration and the mean barrier of the hops that each species encounters together determine whether elemental tracer diffusion is sluggish. A screen of all 57 equiatomic alloys, based on the deviation of the mean kinetically resolved activation barrier from the pure-element average, suggests that the kinetic contribution is sluggish in most of these alloys. This contribution appears to reflect local chemical interactions rather than the number of constituent elements. The eLCE framework in \texttt{pyeCE} provides a computationally tractable, first-principles route to mass transport in compositionally complex materials.
\end{abstract}

\maketitle

\section{Introduction}
Diffusion determines the rate of microstructural evolution and controls phenomena such as phase transformations, creep, oxidation, radiation damage, and ionic conductivity \cite{ardell_trans-interface_2005,darling_extreme_2016,hinrichs_ductile_2025,el-atwani_outstanding_2019,wang_multiscale_2025,van_der_ven_rechargeable_2020}. In compositionally complex materials, such as high-entropy and multi-principal element alloys, the large number of distinct local chemical environments produces a wide distribution of atomic interactions and migration barriers \cite{hsu_clarifying_2024,miracle_critical_2017,miracle_high-entropy_2017,zeng_high-entropy_2022}. Chemical short-range order (SRO) changes how often each of these environments occurs and can alter transport \cite{goiri_role_2019}. Diffusion in high-entropy alloys has also been proposed to be more \emph{sluggish} than expected from their pure constituent elements \cite{yeh_recent_2006,hsu_clarifying_2024}. Sluggish diffusion remains debated because experiments and simulations report both sluggish \cite{tsai_sluggish_2013,daw_sluggish_2021} and non-sluggish or anti-sluggish \cite{zhang_zr_2022,sen_anti-sluggish_2023} behavior. Predicting diffusion in these systems is difficult because thermodynamic and kinetic effects are coupled and both depend on the local chemical environment.

First-principles calculations of vacancy formation energies and migration barriers provide a rigorous basis for modeling diffusion \cite{van_der_ven_vacancy_2010,shang_comprehensive_2016}. However, a disordered alloy contains too many distinct configurations and local environments to compute each one directly with density functional theory (DFT). Cluster expansions \cite{sanchez_generalized_1984} address this problem by predicting the energy of any configuration on a fixed lattice. Local cluster expansions extend this approach to the migration barriers of individual hops \cite{van_der_ven_first-principles_2001}. Kinetic Monte Carlo (kMC) simulations then evolve the system in time by selecting among all possible events according to their rates \cite{bulnes_collective_1998}. The combination of cluster expansions and kMC has been applied to ionic diffusion in electrodes \cite{kaufman_cation_2022,wang_kinetic_2023,puchala_-site_2013,jadidi_kinetics_2023,xiao_kinetic_2018,van_der_ven_nondilute_2008} and electrolytes \cite{morgan_lattice-geometry_2017,deng_fundamental_2022}, vacancy diffusion in concentrated alloys \cite{van_der_ven_first_2005,goiri_role_2019,behara_role_2024}, hydrogen diffusion \cite{samin_first-principles_2019}, and surface diffusion in catalysis \cite{andersen_practical_2019}.

Applying cluster expansions to alloys with many species is challenging because the number of cluster functions grows polynomially with the number of species \cite{muller_constructing_2025}. Cluster expansions of compositionally complex materials therefore contain many coefficients that must be fitted to first-principles data. Neural networks that predict migration barriers directly from the local environment offer an alternative, but they can require large training datasets and do not necessarily satisfy detailed balance \cite{talapatra_best_2025}. The embedded cluster expansion (eCE) uses chemical embedding to reduce the effective dimensionality of the chemical space so that it can be fitted to smaller training sets \cite{muller_constructing_2025}. The embedded local cluster expansion (eLCE) extends this approach to local properties such as the kinetically resolved activation (KRA) barrier of a vacancy hop \cite{lee_diffusion_2026}.

Here, we present an open-source implementation of the eLCE framework coupled to kMC simulations in the \texttt{pyeCE} package. \Cref{sec:methods} reviews the formalism and the workflow. To provide practical guidelines for fitting eLCE models, we show in \cref{sec:benchmark} that models trained on binary and ternary configurations extrapolate to quaternary through senary alloys. Even models with large fitting errors reproduce the qualitative trends in the diffusion quantities. In \cref{sec:applications}, we apply these models to alloys of the group 5 and 6 elements V, Nb, Ta, Cr, Mo, and W. We first show that SRO can change elemental tracer diffusion coefficients quantitatively by shifting the distribution of migration barriers that each element encounters. We then show that the vacancy concentration and the effective migration barrier together determine whether elemental tracer diffusion is sluggish. Finally, we use the deviation of the mean KRA barrier of each element from the average over the pure constituents as a heuristic to screen the kinetic contribution to sluggish diffusion in all equiatomic alloys of these elements.

\section{Methods}
\label{sec:methods}
\subsection{Kinetics of multicomponent materials}
We first summarize the formalism for computing diffusion quantities for both substitutional and interstitial diffusion from first-principles calculations. For a complete description, we refer the reader to several established works \cite{van_der_ven_vacancy_2010,van_der_ven_rechargeable_2020,puchala_casm_2025}. Diffusion is driven by gradients in diffusion potentials $\nabla \tilde{\mu}_{i}$ \cite{kehr_mobility_1989,balluffi_kinetics_2005}. The fluxes of the $c$ mobile species are related to these gradients through the matrix of Onsager transport coefficients $\mathbf{\tilde{L}}$:
\begin{equation}
  \vec{J}=-\frac{1}{\Omega k_B T} \mathbf{\tilde{L}}\nabla \vec{\tilde{\mu}}
  \label{eq:flux}
\end{equation}
where $\Omega$ is the volume per substitutional or interstitial site. For substitutional diffusion, atoms diffuse by exchanging with vacancies and the total number of lattice sites, including vacant sites, is fixed. The diffusion potential is then the difference between the chemical potentials of the species and the vacancy, $\tilde{\mu}_i=\mu_i-\mu_{Va}$. For interstitial diffusion, the diffusion potential is the chemical potential of the diffusing species, $\tilde{\mu}_i=\mu_i$. The elements $\tilde{L}_{ij}$ of the $c\times c$ matrix $\mathbf{\tilde{L}}$ can be obtained from Kubo--Green linear response theory \cite{green_markoff_1954,kubo_statistical-mechanical_1957}:
\begin{equation}
  \tilde{L}_{ij} =
  \frac{\langle
    \sum_{\alpha\in i}\Delta \vec{R}_i^{\alpha}(t) \cdot
  \sum_{\alpha\in j}\Delta \vec{R}_j^{\alpha}(t) \rangle}
  {2dtM}
  \label{eq:kubo_green}
\end{equation}
in which $d$ is the dimensionality of the diffusion network and $M$ is the number of substitutional or interstitial sites in the crystal. The numerator is the ensemble average, denoted by $\langle\cdot\rangle$, of the product of the total displacements of species $i$ and $j$ after time $t$. Each total displacement $\sum_{\alpha\in i}\Delta \vec{R}_i^{\alpha}(t)$ is summed over all atoms $\alpha$ of species $i$.

The displacements $\Delta \vec{R}_i^{\alpha}(t)$ can be sampled from simulated trajectories in which atoms hop to neighboring vacant sites with frequencies given by transition state theory \cite{vineyard_frequency_1957}:
\begin{equation}
  \Gamma=\nu^*\exp\left(\frac{-\Delta E}{k_BT}\right)
  \label{eqn:transition_state}
\end{equation}
where $\nu^*$ is the vibrational prefactor and $\Delta E$ is the migration barrier of the hop. kMC simulations sample such trajectories over time scales far beyond those accessible to molecular dynamics. They require $\nu^*$ and $\Delta E$ for every possible hop.

Estimating $\tilde{L}_{ij}$ from only the initial and final configurations of a trajectory leads to poor statistical convergence. Time-window averaging improves the convergence \cite{binder_applications_1984} by evaluating $\tilde{L}_{ij}$ over all pairs of simulation time steps separated by an interval $\Delta\tau$ and averaging the results.

The flux equations in \cref{eq:flux} can be recast in terms of concentration gradients to obtain Fick's first law:
\begin{equation}
  \vec{J}=-\mathbf{D}\nabla\vec{c}
  \label{eq:flux_concentration}
\end{equation}
where $\mathbf{D}$ is the matrix of intrinsic diffusion coefficients and $\nabla\vec{c}$ is the vector of concentration gradients. Unlike gradients in diffusion potentials, concentration gradients are readily measured in experiments. The matrix $\mathbf{D}$ is related to $\mathbf{\tilde{L}}$ by $\mathbf{D}=\mathbf{\tilde{L}}\mathbf{\tilde{\Theta}}$, where the thermodynamic factor $\mathbf{\tilde{\Theta}}$ characterizes the curvature of the free-energy surface. Methods for computing $\mathbf{\tilde{\Theta}}$ are described in detail elsewhere \cite{lee_first-principles_2026}. The expressions above assume that diffusion is isotropic in $d$ dimensions, as in cubic crystals. In systems with lower symmetry, $\mathbf{\tilde{L}}$ and $\mathbf{D}$ must be evaluated along each crystallographic direction.

Additional metrics characterize other aspects of diffusion. The correlation factor $f_i$ is defined as:
\begin{equation}
  f_i=\frac{\sum_{\alpha \in i} \Delta \vec{R}^\alpha(N^\alpha) ^2}{\sum_{\alpha \in i} N^\alpha \Delta\vec{r}^2}
  \label{eq:correlation_factor}
\end{equation}
where $\Delta \vec{R}^\alpha(N^\alpha)$ is the displacement of atom $\alpha$ of species $i$ after $N^\alpha$ hops and $\Delta\vec{r}^2$ is the square of the elementary hop distance. The correlation factor measures how efficiently successive hops produce net displacement, relative to a random walk. A purely random walk gives $f_i=1$, while $f_i=0$ when each hop reverses the previous one. Simulations of materials with dilute vacancy concentrations, such as metals and alloys, typically contain a single vacancy in the simulation cell. The vacancy correlation factor $f_{Va}$ then converges poorly because only a single vacancy trajectory is sampled. Block averaging improves the convergence \cite{mishin_monte_1997} by dividing the trajectory into blocks containing a fixed number of hops and computing $\Delta\vec{R}$ independently for each block.

The tracer diffusion coefficient can be measured with radiotracer techniques and is among the most commonly reported diffusion quantities. It characterizes the diffusion of tagged atoms in the absence of chemical gradients:
\begin{equation}
  D^*_i=\frac{\langle {\Delta\vec{R}_i^{\alpha}}^{2} \rangle}{2dt}
  \label{eq:tracer}
\end{equation}
Here, the mean-squared displacement (MSD) in the numerator is averaged over the individual atoms $\alpha$ of species $i$. Unlike the numerator of \cref{eq:kubo_green}, it contains no cross-correlations between the displacements of different atoms. In practice, $D^*_i$ is obtained from the slope of the MSD with respect to the simulation time $t$.

In most cases of substitutional diffusion, vacancies are present in dilute concentrations. In the dilute-vacancy limit, quantities such as $\tilde{L}_{ij}$ and $D^*_{i\neq Va}$ scale linearly with the vacancy concentration $x_{Va}$ to leading order \cite{van_der_ven_vacancy_2010}. These quantities can therefore be rescaled by a factor of $x_{Va}^{\mathrm{eqm}}/x_{Va}^{\mathrm{MC}}$, where $x_{Va}^{\mathrm{eqm}}$ is the equilibrium vacancy concentration and $x_{Va}^{\mathrm{MC}}$ is the vacancy concentration in the simulation cell. The equilibrium vacancy concentration can be computed using coarse-graining techniques \cite{belak_effect_2015,li_vacancy_2024,lee_modeling_2026}.

\subsection{Embedded Local Cluster Expansions}
\begin{figure*}
  \centering
  \includegraphics{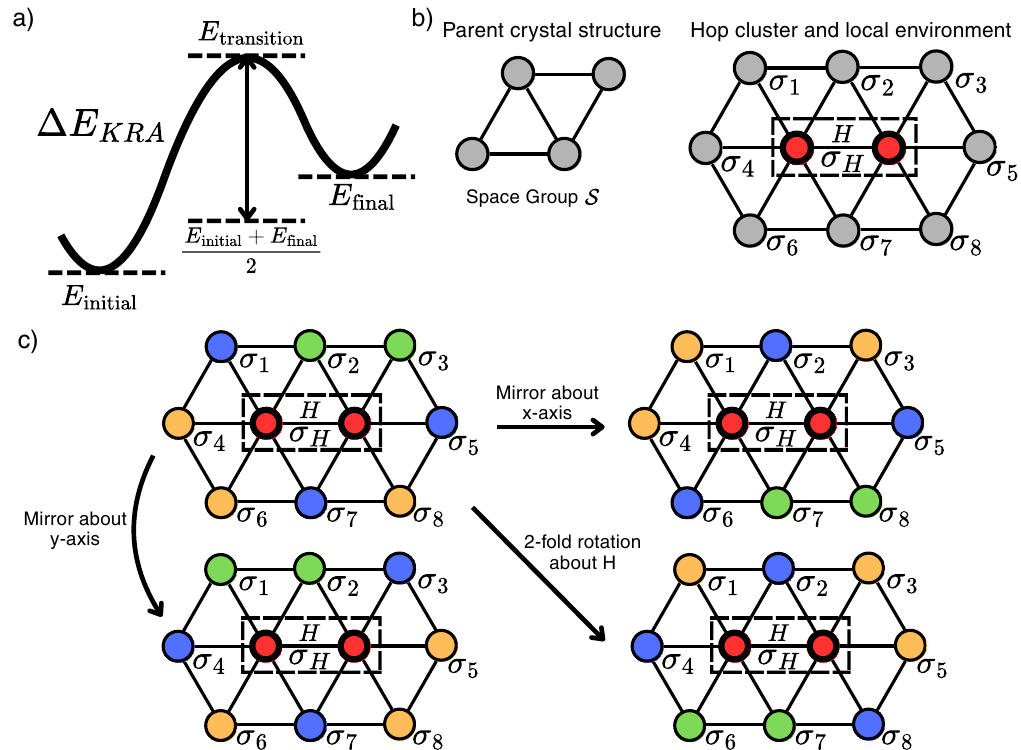}
\caption{a) Definition of the KRA barrier $\Delta E_{KRA}$ (\cref{eqn:kra}). Unlike the migration barrier, $\Delta E_{KRA}$ does not depend on the direction of the hop. b) Local environment of a hop cluster $H$ on a 2D triangular lattice. c) Symmetry operations that leave $H$ invariant, shown on a triangular lattice decorated with different species (colors). $\Delta E_{KRA}$ is invariant under these operations.}
\label{fig:schematic}
\end{figure*}

A migration event involves a single atom exchanging places with a neighboring vacant site. The migration barrier $\Delta E$ in \cref{eqn:transition_state} depends on the migrating species, the local chemical environment of the hop, and the direction of the hop. The forward and backward barriers of a hop differ by the energy difference between its end states. The KRA barrier removes this directional dependence \cite{van_der_ven_first-principles_2001}. It is defined as the difference between the transition-state energy and the average of the two end-state energies:
\begin{equation}
\Delta E_{KRA} = E_{\mathrm{transition}}-\frac{(E_{\mathrm{initial}}+E_{\mathrm{final}})}{2}
\label{eqn:kra}
\end{equation}
as illustrated in \cref{fig:schematic}a. We define the hop cluster $H$ as the pair cluster of the vacancy and the migrating atom. The occupational variable $\sigma_H$ encodes the identity of the migrating atom. $\Delta E_{KRA}$ depends on $\sigma_H$ and on the local ordering vector $\vec{\sigma}_L=[\sigma_1,\sigma_2,\cdots,\sigma_N]$. This vector collects the occupational variables of the $N$ sites within a cutoff distance of the center of $H$, excluding the two sites of $H$ (\cref{fig:schematic}b).

An eLCE expands the configurational dependence of $\Delta E_{KRA}(\vec{\sigma}_L,\sigma_H)$ as \cite{van_der_ven_first-principles_2001,lee_diffusion_2026}:
\begin{equation}
\Delta E_{KRA}(\vec{\sigma}_L,\sigma_H)=K_0+\sum_{\alpha,\beta} K_{\alpha\beta}\Phi_{\alpha\beta}(\vec{\sigma}_L,\sigma_H)
\label{eqn:cluster_expansion}
\end{equation}
Here, $K_0$ is a constant and $K_{\alpha\beta}$ are expansion coefficients fitted to first-principles data. The cluster functions $\Phi_{\alpha\beta}$ are defined as:
\begin{equation}
\label{eqn:cluster_functions}
\Phi_{\alpha\beta}=
\tilde{\psi}_{\beta}(\sigma_H)\prod_{(i,\nu)\in\alpha}\tilde{\phi}_{\nu}(\sigma_i)
\end{equation}
The cluster $\alpha$ is a set of pairs $(i,\nu)$, where $i$ is a site index in the local environment and $\nu$ is a basis-function index. $\tilde{\phi}_{\nu}$ is the projected site basis function with index $\nu$, and $\tilde{\psi}_{\beta}$ is the projected site basis function of $H$ with index $\beta$. In traditional cluster expansions of $c$ species, each site is described by $c$ site basis functions, collected in the vector $\vec{\phi}$. The number of cluster functions therefore grows polynomially with $c$. Embedded cluster expansions limit this growth with embedding matrices $T$ that project the site basis functions onto a lower-dimensional space \cite{muller_constructing_2025}:
\begin{align}
\label{eq:eCE_site_basis_functions}
&\vec{\tilde{\psi}}(\sigma_H) = T\vec{\psi}(\sigma_H) \\
&\vec{\tilde{\phi}}(\sigma_i) = T\vec{\phi}(\sigma_i)
\end{align}
As in traditional cluster expansions, the terms of the local cluster expansion can be interpreted as point, pair, triplet, and higher-order interactions. Without embedding, using the occupational basis for the hop cluster is equivalent to constructing a separate model for each hopping species, as done in previous studies \cite{goiri_role_2019,behara_role_2024,van_der_ven_vacancy_2010,zhang_ab_2022}.

An energy cluster expansion gives the end-state energies $E_{\mathrm{initial}}$ and $E_{\mathrm{final}}$. The forward and backward migration barriers are then recovered from the KRA barrier through \cref{eqn:kra}. When a local cluster expansion for the KRA barrier is combined with an energy cluster expansion, the predicted migration barriers satisfy detailed balance by construction \cite{van_der_ven_first-principles_2001}.

The symmetry of the parent crystal and of the hop cluster reduces the number of independent terms in \cref{eqn:cluster_expansion}. In a local cluster expansion, symmetry equivalence is determined by the subgroup $\mathcal{H} \subseteq \mathcal{S}$ of point-group operations that leave the hop cluster invariant, where $\mathcal{S}$ is the space group of the crystal. Symmetrically equivalent clusters form an \textit{orbit} and share the same expansion coefficient. Because all operations in $\mathcal{H}$ are defined relative to the center of $H$, including $H$ in the cluster functions does not break the symmetry. \Cref{fig:schematic}c illustrates this symmetry equivalence for a triangular lattice. Once all symmetrically distinct orbits are identified, the eLCE coefficients are fitted to first-principles KRA barriers.

\subsection{\texttt{pyeCE} package and general workflow}
The \texttt{pyeCE} package is an open-source \texttt{Python} implementation of the eCE formalism \cite{muller_constructing_2025,muller_pyece_2026}. Documentation and examples for \texttt{pyeCE} are available at \url{https://pyece.readthedocs.io}. In this work, we extend the package to eLCE models of crystalline materials. These models can describe both vacancy-mediated substitutional diffusion in alloys and interstitial diffusion, such as that of Li in electrode materials. \Cref{fig:workflow} illustrates the five steps for building eLCE models, performing kMC simulations, and extracting kinetic properties:
\begin{figure*}
\centering
\includegraphics{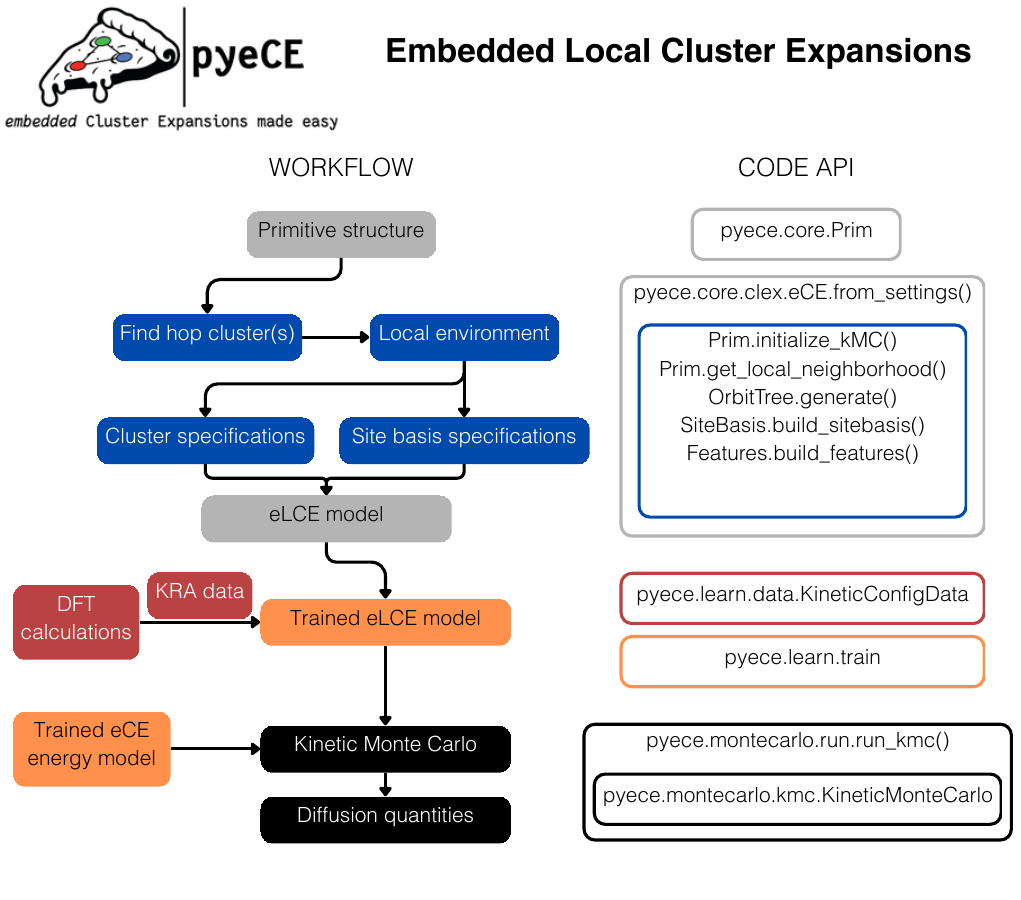}
\caption{Workflow for constructing eLCE models and performing kMC simulations (left), and the \texttt{pyeCE} application programming interface (API) that implements each step (right).}
\label{fig:workflow}
\end{figure*}
\begin{enumerate}
\item \textbf{Initialization.} A user-defined cutoff on the hop distance determines the possible migration pathways. The \texttt{Prim} class defines a hop cluster $H$ for each pathway. The code supports multiple hop types, such as basal and pyramidal hops in hcp or nearest-neighbor and next-nearest-neighbor hops. The space-group symmetry of the crystal generates all hops equivalent to each symmetrically distinct hop. On the bcc lattice, for example, it generates the eight equivalent nearest-neighbor hops from each site. A separate eLCE model is constructed for each symmetrically distinct hop type.
\item \textbf{Enumeration of clusters and initialization of site basis functions.} A cutoff radius from the center of $H$ defines the local environment. The symmetry operation that maps one hop cluster onto an equivalent one also maps their local environments onto each other. Point, pair, triplet, and higher-order clusters are then enumerated within user-defined cutoffs and grouped into orbits under the operations in $\mathcal{H}$. The site basis functions of the hop cluster and of the local environment are constructed separately. The hop cluster and each group of symmetrically equivalent sites in the local environment therefore have their own embedding matrix, with a user-specified embedding dimension. The embedding matrices can be initialized from elemental properties. This initialization has been shown to give accurate eCE energy models \cite{muller_constructing_2025}. The wrapper function \texttt{eCE.from\_settings()} constructs the eLCE model from a single settings file.
\item \textbf{Model fitting.} Each training data point consists of a first-principles KRA barrier and the initial and final configurations of the hop. The code identifies the hop cluster from these configurations and stores the data point in a \texttt{KineticConfigData} object. The cluster functions evaluated for each data point are the inputs to a linear model or to a nonlinear model such as a neural network. The \texttt{train} module optimizes the model parameters and the embedding matrices by gradient descent on a loss function, with optional L1 and L2 regularization to reduce overfitting.
\item \textbf{Kinetic Monte Carlo.} Rejection-free kMC simulations give access to diffusion at experimentally relevant timescales. The simulations evaluate the KRA barriers with the fitted eLCE models and the end-state energies with an eCE model. Initial configurations are drawn from canonical Monte Carlo at the temperature of interest. At each kMC step, all possible hops are enumerated and the migration barrier of each hop is computed as $\Delta E=\Delta E_{KRA}+(E_{\mathrm{final}}-E_{\mathrm{initial}})/2$ (\cref{eqn:kra}). A single hop is then selected with a probability proportional to its rate (\cref{eqn:transition_state}) \cite{bulnes_collective_1998}. The simulation time is advanced by $\Delta t=-\ln(\zeta)/\sum_i \Gamma_i$, where $\zeta$ is a uniformly distributed random number in $(0,1]$ and $\sum_i \Gamma_i$ is the total rate of all possible hops. Converged diffusion quantities require an ensemble of independent trajectories, each run for a specified number of kMC steps. The \texttt{KineticMonteCarlo} class runs these trajectories in parallel on GPUs and uses \texttt{numba} for just-in-time compilation. The driver function \texttt{run\_kmc()} creates and runs a \texttt{KineticMonteCarlo} instance.

\item \textbf{Extraction of kinetic properties.} The code records the displacement of every atom at each kMC step. From these displacements, it computes $\tilde{L}_{ij}$, $f_{i}$, and $D_i^*$ through \cref{eq:kubo_green,eq:correlation_factor,eq:tracer}, using time-window averaging and block averaging.
\end{enumerate}

\subsection{Training data and simulation settings}
To demonstrate the capabilities of the eLCE framework, we use a previously reported DFT dataset of KRA barriers for vacancy hops on the bcc lattice \cite{lee_diffusion_2026}. The dataset covers binary through senary alloys of the group 5 and 6 elements V, Nb, Ta, Cr, Mo, and W. The local environment of each eLCE model contained all sites within 8\,\AA{} of the center of the hop cluster, and pair clusters were included up to a length of 5\,\AA{}. Augmenting each cluster with the hop cluster extended point clusters to pair clusters and pair clusters to triplet clusters. The resulting cluster set comprised 1 empty cluster, 21 point clusters, 26 pair clusters, and 7 triplet clusters. This set includes clusters both with and without the hop cluster. Occupational and Chebyshev site basis functions were used for the hop cluster and the local environment, respectively. These basis functions were projected onto a lower-dimensional space by trainable embedding matrices. The embedding matrices were initialized from elemental properties \cite{muller_constructing_2025} and renormalized after each training iteration. Symmetrized cluster functions built from tensor products of the embedded basis functions were the inputs to a 4-layer neural network (32 $\times$ 32 $\times$ 8 $\times$ 1) with ReLU activations in the hidden layers and a linear output layer. The reference dataset was split into training, validation, and test sets containing 80\%, 10\%, and 10\% of the data, respectively. The training loss was minimized using the Adam optimizer with a learning rate scheduler. Overfitting was controlled through early stopping on the validation set and L1/L2 regularization.

The end-state energies in the kMC simulations were evaluated with a previously parametrized eCE energy model of embedding dimension 3 \cite{lee_diffusion_2026}. All kMC simulations used a single vacancy in a $10\times10\times10$ conventional bcc supercell (2000 sites) and a vibrational prefactor of $\nu^*=1\times10^{12}$\,Hz. Each simulation consisted of 500 independent trajectories of 50,000 steps each (25 million steps in total). Each kMC step required approximately $5\times10^{-4}$\,s on an NVIDIA L40S GPU. This cost is comparable to that of other kMC implementations \cite{hoffmann_kmos_2014,deng_kmcpy_2023}. A key advantage of our implementation is its ability to execute multiple independent kMC trajectories in parallel on the same GPU, thus improving the overall computational throughput.

\section{Results}
\label{sec:results}
\subsection{Benchmarking eLCE models}
\label{sec:benchmark}
\subsubsection{Data efficiency from chemical extrapolation}
\begin{figure}[h!]
\centering
\includegraphics{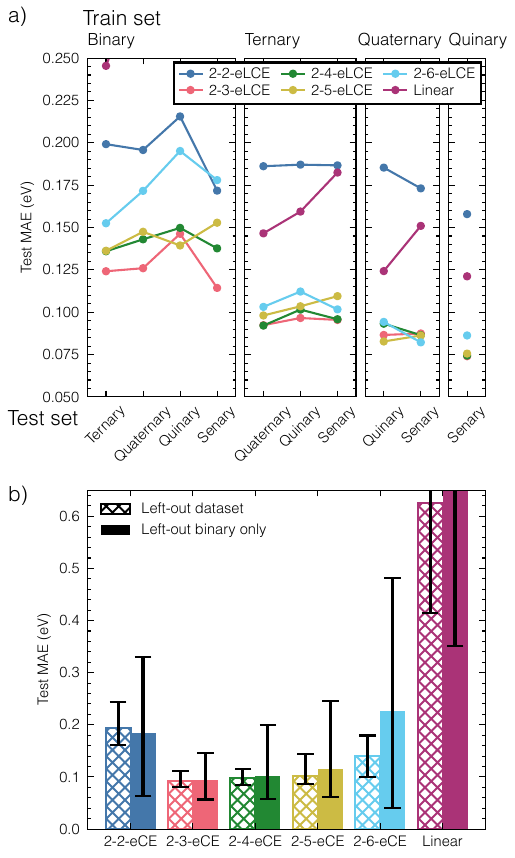}
\caption{a) Test MAE of $n$-$k$-eLCE models and a linear model trained on datasets of increasing compositional complexity. The embedding dimensions of the hop cluster and the local environment are $n$ and $k$, respectively. Each panel corresponds to a training set that contains configurations with at most the number of elements indicated at the top. The test set contains all remaining configurations, which have more elements, as indicated at the bottom. For example, the quinary training set contains all binary, ternary, quaternary, and quinary data, and the model is tested on senary data. b) Test MAE from leave-one-pair-out cross-validation. For each pair of elements, the test set contains all configurations with both elements of the pair. Bars show the test MAE averaged over all excluded pairs. Error bars show the minimum and maximum test MAE across these pairs. Hatched bars show the error on all excluded configurations that contain a given pair. Solid bars show the error on the excluded configurations that contain only that pair.}
\label{fig:extrapolation}
\end{figure}

DFT calculations of transition-state energies with the nudged elastic band method \cite{henkelman_climbing_2000,henkelman_improved_2000} are computationally expensive. More efficient workflows reduce this cost by combining universal machine-learned interatomic potentials with static DFT calculations \cite{lee_diffusion_2026}. However, sampling the entire chemical space of a compositionally complex alloy remains computationally prohibitive even with these workflows. In this section, we benchmark the data efficiency of eLCE models with respect to the chemical complexity of the training data and the coverage of element pairs in the training data.

In an $n$-$k$-eLCE model, $n$ is the embedding dimension of the hop cluster and $k$ is the embedding dimension of the local environment. \Cref{fig:extrapolation}a shows the test mean absolute error (MAE) of $n$-$k$-eLCE models trained on datasets of increasing compositional complexity. Each model was trained on configurations with up to a given number of elements and tested on all configurations with more elements. We fix $n=2$ and vary $k$ from 2 to 6. For comparison, we also include a linear model without embedding that was trained by ridge regression.

When trained only on binary data, the 2-3-eLCE model has extrapolation errors of 0.1\,eV to 0.15\,eV (\cref{fig:extrapolation}a). Adding ternary data to the training set lowers the errors of all models with $k=3$--6 from as much as 0.2\,eV to approximately 0.1\,eV on every test set. In contrast, the 2-2-eLCE model and the linear model extrapolate poorly and have errors of 0.15\,eV to 0.2\,eV. These results suggest that, as reported for eCE energy models \cite{muller_constructing_2025}, the KRA barriers in the senary composition space can be described by binary and ternary interactions.

In a leave-one-pair-out cross-validation, all configurations containing both elements of a given pair were excluded from the training set and used as the test set. \Cref{fig:extrapolation}b summarizes the cross-validation results for all models and all excluded pairs. For $k=3$--5, the MAE averaged over the excluded pairs is approximately 0.1\,eV. On configurations that contain only the excluded binary pair, these models have comparable average errors but a larger spread. The 2-2-eLCE model has larger errors of approximately 0.2\,eV, and the 2-6-eLCE model has errors of 0.14\,eV on all excluded configurations and 0.23\,eV on the binary configurations alone. The errors of the linear model are the largest and range from 0.6\,eV to 1\,eV. These results indicate that eLCE models with $k=3$--5 generalize to combinations of elements that are absent from the training set.

Of the models in \cref{fig:extrapolation}, the 2-3-eLCE model extrapolates best in chemical space when trained on binary data alone. We attribute this performance to two features of the model. First, the local cluster expansion exploits the symmetry of the hop cluster. This symmetry reduces the number of independent coefficients in every model of \cref{fig:extrapolation}, including the linear model, and a model without it would require more coefficients and thus more training data. Second, chemical embedding captures the similarities between chemical species. Only the eLCE models benefit from this second feature, and it allows them to extrapolate from binary data to configurations with more elements. When trained on 80\% of the full dataset (1,945 data points), the 2-3-eLCE model reaches a test MAE of 70\,meV. This training set is orders of magnitude smaller than the datasets typically required to train neural network models \cite{xu_revealing_2022,huang_machine_2023,xing_vacancy_2024,xing_neural_2024}. We use this model, labeled eLCE-70, in the next section.

\subsubsection{Impact of fitting error on diffusion}
\begin{figure}[h!]
\centering
\includegraphics{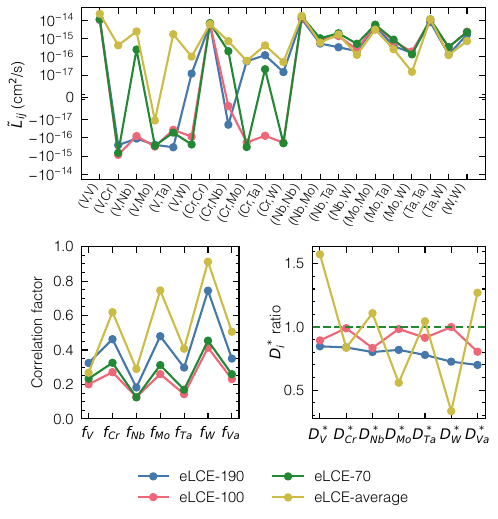}
\caption{Diffusion quantities computed with eLCE models of different accuracy for the equiatomic senary alloy. Each model is labeled eLCE-$X$, where $X$ is its test MAE in meV. The eLCE-average model assigns each hopping species a constant KRA barrier equal to its mean value in the training set. Top: Onsager transport coefficients $\tilde{L}_{ij}$ of the four models. Bottom left: correlation factors $f_i$ of the six elements and the vacancy. Bottom right: tracer diffusion coefficient $D^*_i$ of each model as a ratio to that of the most accurate model (eLCE-70).}
\label{fig:error_comparison}
\end{figure}
The effect of the fitting error of an eLCE model on the predicted diffusion quantities has not been quantified. To assess this effect, we performed kMC simulations of the equiatomic senary alloy with eLCE models of different accuracy and computed $\tilde{L}_{ij}$, $f_i$, and $D^*_i$. These simulations were performed at a temperature of 2500~K where the alloy is in the disordered state \cite{lee_diffusion_2026}. In addition to the eLCE-70 model, we use two models with test MAEs of 190\,meV and 100\,meV, labeled eLCE-190 and eLCE-100. These models were constructed using different embedding dimensions and smaller datasets. We also consider a model, labeled eLCE-average, that assigns each hopping species a constant KRA barrier equal to its mean value in the training set.

\Cref{fig:error_comparison} shows that all four models predict diagonal Onsager transport coefficients of the same order of magnitude. The off-diagonal coefficients differ more between the models, in some cases in sign. However, the off-diagonal coefficients are generally smaller than the diagonal coefficients because the kinetic coupling between different species is weak. These small quantities are difficult to converge numerically, and differences in their values should be interpreted with caution. The correlation factors of the eLCE-100 and eLCE-70 models agree to within 0.06 for all species. The eLCE-190 model gives correlation factors that are systematically higher than those of the eLCE-70 model by up to 0.3, while those of the eLCE-average model differ by nearly 0.5. The trends in the correlation factors across species are similar for all four models.

Similar to the correlation factors, the tracer diffusion coefficients of the eLCE-average model deviate most from those of the eLCE-70 model (\cref{fig:error_comparison}). The eLCE-average model overpredicts the tracer diffusion coefficients of V, Nb, Ta, and Va and underpredicts those of Cr, Mo, and W. Its largest deviation is a threefold underestimate of $D^*_W$. The eLCE-average model assigns the same KRA barrier to every local environment of a given hopping species. This assignment removes the distribution of KRA barriers that shapes the diffusion pathways and correlations in compositionally complex alloys. In contrast, the eLCE-190 and eLCE-100 models give tracer diffusion coefficients that are 0.70 to 0.85 and 0.8 to 1.0 times those of the eLCE-70 model, respectively. Both models underpredict or match the tracer diffusion coefficients of the eLCE-70 model for every species. These results indicate that models with test MAEs as large as 190\,meV reproduce the qualitative trends in the diffusion quantities.

\subsection{Diffusion in compositionally complex alloys}
\label{sec:applications}
Having benchmarked the data efficiency of eLCE models and the effect of their fitting error on the predicted diffusion quantities, we now use the eLCE-70 model to study diffusion in alloys of the group 5 and 6 elements. We examine the effect of SRO on diffusion, metrics for sluggish diffusion, and the screening of equiatomic alloys for sluggish diffusion.

\subsubsection{Impact of SRO on diffusion}
\begin{figure*}
\centering
\includegraphics{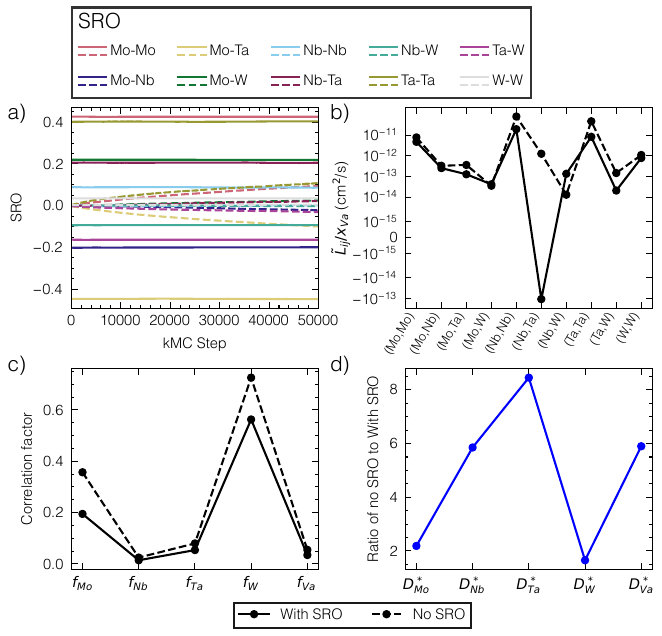}
\caption{kMC simulations of equiatomic NbMoTaW at 1000\,K starting from equilibrium and random configurations. \textbf{a)} Warren--Cowley SRO parameters of first nearest-neighbor pairs as a function of kMC step. \textbf{b)} Onsager transport coefficients divided by the vacancy concentration $x_{Va}$, \textbf{c)} correlation factors, and \textbf{d)} tracer diffusion coefficients of the equilibrium and random alloys. In (a)--(c), solid and dashed lines represent the equilibrium and random alloys, respectively. In (d), the ratio of the tracer diffusion coefficients of the random alloy to those of the equilibrium alloy is plotted.}
\label{fig:sro_plot}
\end{figure*}

The effects of SRO on the properties of compositionally complex alloys, including diffusion \cite{goiri_role_2019}, have been widely discussed. We investigate the effect of SRO on diffusion with kMC simulations of equiatomic NbMoTaW at 1000\,K. At this temperature, the alloy remains a disordered solid solution but develops appreciable SRO (\cref{fig:sro_plot}a). We ran two sets of kMC simulations, with initial configurations generated either by canonical Monte Carlo at 1000\,K or by random assignment of species to lattice sites. We refer to these as the equilibrium alloy and the random alloy, respectively.

\Cref{fig:sro_plot}a shows the Warren--Cowley SRO parameters of the first nearest-neighbor pairs during the simulations. The SRO parameters of the equilibrium alloy remain constant, whereas those of the initially random alloy evolve from zero toward the equilibrium values. The random alloy has not reached equilibrium after 50,000 steps and provides a non-equilibrium reference. The Mo--Ta and Mo--Nb pairs have large negative SRO parameters in the equilibrium alloy ($-0.45$ and $-0.20$, respectively). These pairs occur more often than in a random alloy.

\Cref{fig:sro_plot}b--d compare $\tilde{L}_{ij}$, $f_i$, and $D_i^*$ of the equilibrium and random alloys. Quantities that depend on the vacancy concentration were rescaled to remove this dependence. The random alloy has larger $\tilde{L}_{Nb,Nb}$ and $\tilde{L}_{Ta,Ta}$ and generally larger off-diagonal coefficients than the equilibrium alloy. The off-diagonal coefficient $\tilde{L}_{Nb,Ta}$ also changes sign, from positive in the random alloy to negative in the equilibrium alloy. The correlation factors are higher in the random alloy for all species. The difference is largest for $f_{Mo}$ and $f_{W}$, at approximately 0.15. The tracer diffusion coefficients are also higher in the random alloy for all species. The ratio between the two alloys is largest for $D^*_{Ta}$, followed by $D^*_{Nb}$ and $D^*_{Va}$, and smallest for $D^*_{Mo}$ and $D^*_{W}$.

\begin{figure}[h!]
\centering
\includegraphics{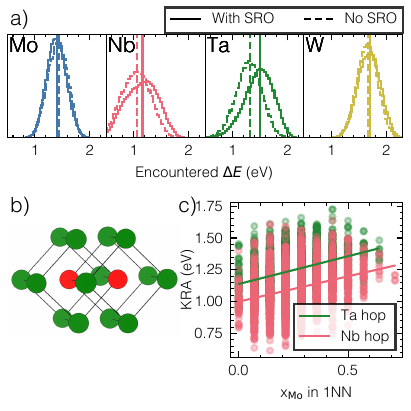}
\caption{\textbf{a)} Distributions of the migration barriers ($\Delta E$ in \cref{eqn:transition_state}) encountered by each hopping element during the kMC simulations of the equilibrium and random alloys. During each kMC step, a vacancy encounters 8 possible hops, which contribute to the distribution. Vertical lines mark the mean of each distribution. \textbf{b)} First nearest-neighbor (1NN) shell (green) of the hop cluster (red) in the bcc lattice. \textbf{c)} KRA barriers of Nb and Ta hops as a function of the Mo concentration in the 1NN shell of the hop cluster. Lines show linear fits for each hopping species.}
\label{fig:sro_plot_2}
\end{figure}

The origin of these differences can be understood from the distributions of migration barriers encountered by each species during the simulations (\cref{fig:sro_plot_2}a). For Nb and Ta, the mean of these distributions is approximately 0.1\,eV and 0.2\,eV higher in the equilibrium alloy with SRO. The distributions of Mo and W are nearly unchanged. The higher barriers slow the hops of Nb and Ta and reduce $\tilde{L}_{Nb,Nb}$, $\tilde{L}_{Ta,Ta}$, $D^*_{Nb}$, and $D^*_{Ta}$. This shift can be rationalized in terms of the local chemical environment of the hop. The equilibrium alloy contains more Nb--Mo and Ta--Mo nearest-neighbor pairs than the random alloy. The first nearest-neighbor (1NN) shell of the hop cluster (\cref{fig:sro_plot_2}b) therefore contains more Mo when Nb or Ta hops. To quantify this effect, we computed the KRA barriers of Nb and Ta hops as a function of the Mo concentration in the 1NN shell (\cref{fig:sro_plot_2}c). According to the linear fits, the KRA barrier increases by approximately 0.3\,eV as the Mo concentration in this shell increases from 0 to 0.7. Mo neighbors raise the typical KRA barriers of Nb and Ta hops. A similar trend is found when W replaces Mo in this analysis.

\subsubsection{Metrics for sluggish diffusion}
Sluggish diffusion in multi-principal element alloys remains debated, in part because intrinsic diffusion coefficients are difficult to measure in compositionally complex systems. A proposed benchmark compares the vacancy tracer diffusion coefficient $D^*_{Va}$ of an alloy with the composition-weighted geometric mean of the pure-element values \cite{daw_sluggish_2021}. This geometric mean is the rule-of-mixtures (ROM) estimate obtained by interpolating the vacancy migration barriers linearly between the pure constituents. The ratio of $D^*_{Va}$ to the ROM estimate is the enhancement factor for the vacancy tracer diffusion coefficient. Values below one indicate sluggish diffusion, and values above one indicate anti-sluggish diffusion. With this metric, the position of the vacancy KRA distribution relative to the ROM estimate was found to predict whether vacancy diffusion in an alloy is sluggish, while thermodynamic effects were comparatively minor \cite{lee_diffusion_2026}.

The vacancy tracer diffusion coefficient characterizes the fastest diffusion mode in an alloy but cannot be measured experimentally. It also contains no information about the slower modes that are limited primarily by the vacancy concentration \cite{lee_first-principles_2026}. Elemental tracer diffusion coefficients can be measured with the radiotracer method and have revealed anti-sluggish diffusion in several systems \cite{sen_anti-sluggish_2023,zhang_zr_2022}. Here, we extend the enhancement factor to the tracer diffusion of individual elements to investigate the origins of sluggish diffusion.

As in a pure element, the elemental tracer diffusion coefficient in a concentrated alloy can be written as the product of three contributions:
\begin{align}
D^*_i &= x_{Va}(\vec{x},T)\, f_i(\vec{x},T)\, D_0 \exp \left(-\frac{\Delta G^{\mathrm{mig}}_i(\vec{x},T)}{k_B T} \right) \\
&= x_{Va}(\vec{x},T)\, f_i(\vec{x},T)\, \hat{D}^*_i
\end{align}
Here, $\vec{x}$ is the alloy composition, and $x_{Va}$ and $f_i$ are the vacancy concentration and the correlation factor. The prefactor is $D_0 = Z {\Delta\vec{r}}^2\nu^*/6$, where $Z$ is the coordination number and $\Delta\vec{r}$ is the hop distance. The effective migration free energy of species $i$ for tracer diffusion is $\Delta G^{\mathrm{mig}}_i$. The quantity $\hat{D}^*_i$ excludes the vacancy concentration and the correlation factor and can be obtained directly from kMC simulations. The enhancement factor $\eta_i$ of species $i$ then factorizes into three contributions:
\begin{align}
\eta_i &= \frac{D^*_{i,\mathrm{alloy}}}{\langle D^*_{i,\mathrm{pure}} \rangle} \\
&= \eta^{Va}
\eta^{f}_{i}
\eta^{D}_{i}
= \frac{x_{Va}}{\langle x_{Va} \rangle}
\frac{f_i}{\langle f_i \rangle}
\frac{\hat{D}^*_i}{\langle \hat{D}^*_i \rangle}
\end{align}
Here, $\eta^{Va}$, $\eta^{f}_{i}$, and $\eta^{D}_{i}$ are the enhancements of the vacancy concentration, the correlation factor, and the migration term $\hat{D}^*_i$ relative to the ROM reference. The angular brackets $\langle \cdot \rangle$ denote a composition-weighted geometric mean over the pure constituents of the alloy. The vacancy concentration typically follows Arrhenius behavior \cite{lee_modeling_2026}, and $\hat{D}^*_i$ depends exponentially on the migration barrier. The geometric mean therefore corresponds to a linear interpolation of the vacancy formation energies and migration barriers between the pure elements. The reference $\langle D^*_{i,\mathrm{pure}} \rangle$ is the geometric mean of the dilute tracer diffusion coefficients of species $i$ in each pure constituent of the alloy.

\begin{figure}[h!]
\centering
\includegraphics{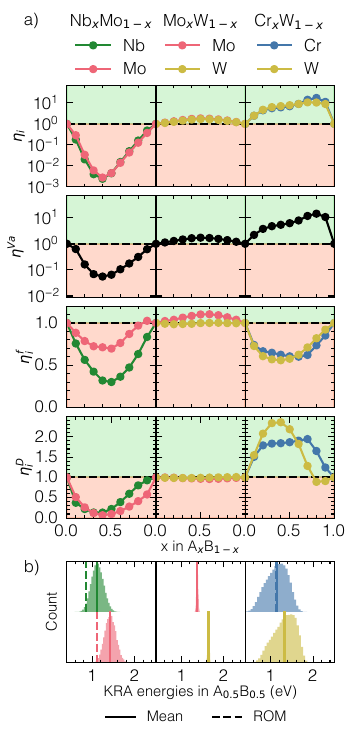}
\caption{\textbf{a)} Enhancement factors $\eta_{i}$, $\eta^{Va}$, $\eta^{f}_{i}$, and $\eta^{D}_{i}$ of Nb$_x$Mo$_{1-x}$, Mo$_x$W$_{1-x}$, and Cr$_x$W$_{1-x}$ as a function of composition $x$. The NbMo and MoW alloys were simulated at 1500\,K and the CrW alloy at 2500\,K. Green and red shading indicate anti-sluggish ($\eta > 1$) and sluggish ($\eta < 1$) regimes, respectively. Dashed lines show the ROM reference. \textbf{b)} KRA barrier distributions of each hopping species at the equiatomic composition. Dashed lines mark the ROM estimate and solid lines mark the mean of each distribution.}
\label{fig:alloy_enhancement}
\end{figure}

We computed $\eta_{i}$, $\eta^{Va}$, $\eta^{f}_{i}$, and $\eta^{D}_{i}$ for the binary alloys Nb$_x$Mo$_{1-x}$, Mo$_x$W$_{1-x}$, and Cr$_x$W$_{1-x}$ across a range of compositions $x$ (\cref{fig:alloy_enhancement}a). The NbMo and MoW alloys were simulated at 1500\,K and the CrW alloy at 2500\,K, so that the NbMo and MoW alloys remain above their order-disorder temperatures and the CrW alloy remains above its miscibility gap \cite{lee_modeling_2026,lee_diffusion_2026}. Vacancy diffusion has previously been identified as sluggish in NbMo, anti-sluggish in CrW, and anti-sluggish above the percolation threshold in MoW \cite{lee_diffusion_2026}. The elemental enhancement factors $\eta_i$ in NbMo fall as low as about $10^{-3}$ and confirm that tracer diffusion of both elements is sluggish. In contrast, elemental tracer diffusion in MoW and CrW is anti-sluggish across the entire composition range. The enhancement factors are largest in CrW, where $\eta_i$ approaches 10.

The vacancy contribution $\eta^{Va}$ follows the same trends as $\eta_i$ (\cref{fig:alloy_enhancement}a). It falls below $10^{-1}$ in NbMo, remains close to 1 in MoW, and rises above 10 in CrW. These trends can be rationalized qualitatively in terms of bonding energetics. Forming a vacancy breaks bonds, so strong bonding between the constituent elements lowers the vacancy concentration, whereas weak or unfavorable bonding raises it \cite{lee_modeling_2026}. In NbMo, strong Nb--Mo bonding lowers the vacancy concentration, while unfavorable Cr--W bonding raises it in CrW.

The kinetic contribution $\eta^{D}_{i}$ to the enhancement factor can be understood by comparing the KRA barrier distributions at the equiatomic composition with the ROM estimate (\cref{fig:alloy_enhancement}b). In NbMo, the KRA distributions of both Nb and Mo lie above the ROM value and reduce $\eta^{D}_{Nb}$ and $\eta^{D}_{Mo}$ well below unity. In MoW, the KRA distributions are narrow and centered on the ROM value, and $\eta^{D}_{Mo}$ and $\eta^{D}_{W}$ remain close to unity across all compositions. Although the means of the KRA distributions in CrW also lie close to the ROM estimate, $\eta^{D}_{Cr}$ and $\eta^{D}_{W}$ reach values of up to about 2. This observation suggests that $\eta^{D}_{i}$ is controlled by the shape of the KRA distribution rather than by its mean when the mean lies close to the ROM estimate. In this case, the hops in the low-energy tail of a broad KRA distribution enhance mobility and lower $\Delta G^{\mathrm{mig}}_i$.

\begin{figure}[h!]
\centering
\includegraphics{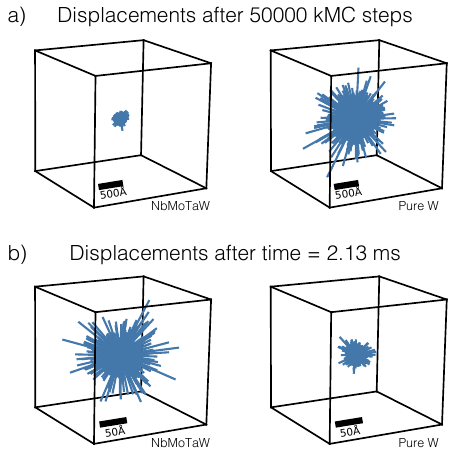}
\caption{Displacement of the vacancy in each trajectory after \textbf{a)} 50,000 kMC steps and \textbf{b)} a simulated time of 2.13\,ms in NbMoTaW (left) and pure W (right) at 1000\,K. Black bars are scale bars.}
\label{fig:trajectories}
\end{figure}

Correlation effects produce fragmented diffusion pathways in alloys \cite{xing_vacancy_2024,chattopadhyay_larger_2026,xing_fragmenting_2026}. The factor $\eta^{f}_{i}$ quantifies the effect of these correlations on sluggish diffusion. It deviates from unity in NbMo and CrW but remains close to unity in MoW. This difference is consistent with the broad KRA distributions of NbMo and CrW and the narrow distributions of MoW (\cref{fig:alloy_enhancement}b). While strong correlations reduce the mean-squared displacement per hop, they do not necessarily slow diffusion, as the time per kMC step depends exponentially on the migration barriers of the available hops. \Cref{fig:trajectories} illustrates this point for NbMoTaW and pure W at 1000\,K. At this temperature, NbMoTaW and pure W have vacancy correlation factors of $f_{Va}=0.035$ and $f_{Va}=1$, respectively. After 50,000 kMC steps, the vacancy in NbMoTaW has explored a smaller region than the vacancy in pure W (\cref{fig:trajectories}a). Pure W, however, has the highest vacancy migration barrier of the six elements considered here and a much lower hop frequency. At a fixed simulated time of 2.13\,ms, the vacancy in NbMoTaW has explored the larger region (\cref{fig:trajectories}b). The higher hop frequencies in NbMoTaW outweigh its stronger correlations and give the vacancy a higher tracer diffusion coefficient than in pure W. Hop statistics alone are therefore not a reliable measure of diffusion kinetics.

\subsubsection{High-throughput screening of alloys}
\begin{figure}[h!]
\centering
\includegraphics{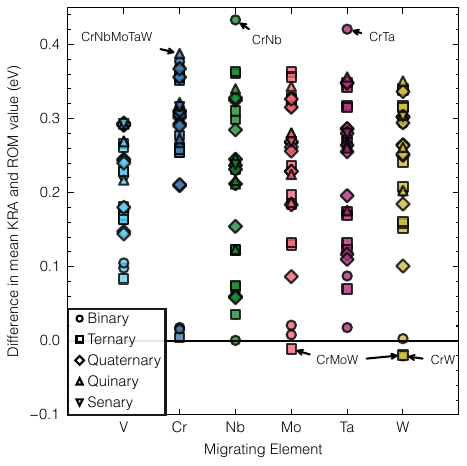}
\caption{Deviation of the mean KRA barrier of each migrating element from its ROM value for all equiatomic alloys of V, Nb, Ta, Cr, Mo, and W. Marker shapes indicate the number of elements in the alloy. The alloys with the three largest positive deviations and all alloys with negative deviations are labeled.}
\label{fig:screening}
\end{figure}

The results above show that the sluggishness of elemental tracer diffusion originates primarily from two sources: a vacancy concentration that is lower or higher than the ROM estimate, and KRA barriers that deviate from their ROM values. When these deviations are large, diffusion can be clearly sluggish or anti-sluggish. When they are small, as for $\eta^{D}_{i}$ in MoW and CrW, kMC simulations are necessary to determine whether diffusion is sluggish. Here, we screen the kinetic contribution $\eta^{D}_{i}$ across all 57 equiatomic alloys of the six elements. The screening heuristic is the deviation of the mean KRA barrier of each species from its ROM value. The vacancy contribution $\eta^{Va}$ is not included in this screen but can be estimated qualitatively from phase diagrams \cite{lee_modeling_2026}.

\Cref{fig:screening} shows the result of this screen. Only Mo in CrMoW and W in CrMoW and CrW have mean KRA barriers slightly below the ROM estimate. These alloys show a weak tendency toward anti-sluggish diffusion. All other deviations are positive and indicate a tendency toward sluggish diffusion. These results suggest that the kinetic contribution to elemental tracer diffusion in this compositional space is largely sluggish, as previously observed for vacancy diffusion \cite{lee_diffusion_2026}. We find no clear correlation between compositional complexity and the deviation from the ROM value. The largest positive deviations occur in the binary alloys CrNb and CrTa. This observation suggests that local chemical interactions, rather than the number of constituent elements, govern the kinetic contribution to sluggish diffusion.

\section{Discussion}
\label{sec:discussion}
Modeling bulk diffusion in compositionally complex alloys has been a long-standing challenge. Atomistic simulation methods that can treat the many local environments in these alloys have been lacking, which has limited our understanding of the mechanisms that govern atomic transport. We implemented the eLCE framework in the \texttt{pyeCE} package and coupled it to kMC simulations that reach experimentally relevant timescales. The implementation supports both vacancy-mediated substitutional diffusion and interstitial diffusion in bulk crystalline materials. Chemical embedding reduces the amount of training data that eLCE models require in compositionally complex systems. Trained only on binary and ternary configurations, a 2-3-eLCE model extrapolates to quaternary through senary alloys with an error of about 0.1\,eV. Models with test MAEs as large as 190\,meV also reproduce the qualitative trends in the predicted diffusion quantities. Together, these features make predictive diffusion modeling in compositionally complex alloys computationally tractable. We used this capability to examine the effect of SRO on diffusion, to quantify sluggish diffusion, and to screen all equiatomic alloys of the group 5 and 6 elements for the kinetic contribution to sluggish diffusion.

\Cref{fig:sro_plot,fig:sro_plot_2,fig:alloy_enhancement} show that the mean barrier of the hops that a diffusing species encounters is a primary factor governing diffusion in these alloys. This dependence was previously established for the vacancy tracer diffusion coefficient \cite{lee_diffusion_2026}. Our results show that it also extends to the tracer diffusion coefficients of the constituent elements and to the effect of SRO. SRO affects diffusion by changing the local chemical environment of each hop and with it the distribution of migration barriers. In NbMoTaW at 1000\,K, the mean migration barriers of Nb and Ta are higher in the equilibrium alloy, and $D^*_{Nb}$ and $D^*_{Ta}$ are lower than in the random alloy by factors of about 6 and 8, respectively. Because the random alloy partially orders during the simulations (\cref{fig:sro_plot}a), its diffusion coefficients are estimates rather than exact values for a perfectly disordered alloy. These factors are therefore likely lower bounds on the effect of SRO.

Combining eLCE models for the KRA barriers with an eCE model for the vacancy thermodynamics and end-state energies allows us to separate the vacancy, kinetic, and correlation contributions to the enhancement factor and thereby identify the origin of sluggish or anti-sluggish diffusion. In NbMo, $\eta^{Va}$ and $\eta^{D}_{i}$ are both well below unity and contribute comparably to the low $\eta_i$ of about $10^{-3}$ (\cref{fig:alloy_enhancement}a). In contrast, $\eta^{Va}$ exceeds $\eta^{D}_{i}$ in MoW and CrW and is the main source of their anti-sluggish diffusion. These results suggest that bonding controls elemental tracer diffusion through both the vacancy formation energy and the KRA barriers. This conclusion differs from that for vacancy tracer diffusion, because $D^*_{Va}$ does not depend on the vacancy concentration and is instead dominated by the KRA distribution \cite{lee_diffusion_2026}. The correlation contribution $\eta^{f}_{i}$ modifies $\eta_i$ less than $\eta^{Va}$ does. In CrW, the means of the broad KRA distributions lie close to the ROM estimate, yet these distributions raise $\eta^{D}_{i}$ to about 2 and lower $\eta^{f}_{i}$ to about 0.6, so that the two contributions partially offset each other. The mean KRA barrier thus determines the kinetic contribution only when it deviates clearly from the ROM estimate, as in NbMo. In this case, the deviation of the mean KRA barrier from the ROM value (\cref{fig:screening}) provides a first screening step for alloy design. Alloys with large deviations can be identified as candidates for a sluggish or anti-sluggish kinetic contribution without kMC simulations. For alloys with small deviations, such as CrW, kMC simulations with eLCE models remain necessary. Because the vacancy contribution can outweigh the kinetic contribution, as in CrW, the screen should be combined with an estimate of $\eta^{Va}$.

The screen of all equiatomic alloys of the six elements suggests that the kinetic contribution to elemental tracer diffusion is sluggish in most of these alloys (\cref{fig:screening}). Because this contribution does not correlate with the number of constituent elements, it appears to reflect local chemical interactions rather than compositional complexity. Combined with the enhancement factors of NbMo, MoW, and CrW, these results indicate that two quantities govern diffusion in these alloys: the vacancy concentration and the mean barrier of the hops that each species encounters. Understanding why the mean KRA barriers deviate from the ROM estimate is therefore central to understanding sluggish diffusion. Local lattice distortions do not account for these deviations \cite{lee_diffusion_2026}, and our results instead point to local bonding. Whether atomic size or another property of the constituent elements controls this bonding contribution remains an open question. The present analysis is also limited to the group 5 and 6 elements on the bcc lattice. Whether the vacancy concentration and the mean KRA barrier remain the dominant factors in other compositionally complex alloys and crystal structures warrants further investigation. The eLCE framework in \texttt{pyeCE} provides a computationally tractable, first-principles route to these questions and, more broadly, to mass transport in compositionally complex materials.

\section{Data availability}
The data used in the present work are available through the Materials Cloud database [\url{https://archive.materialscloud.org/record/2026.191}].

\section{Code availability}
The source code of \texttt{pyeCE} can be found at \url{https://github.com/epfl-mades/pyece} with its documentation at \url{https://pyece.readthedocs.io/en/latest/}.

\section{Acknowledgments}
We acknowledge support from NCCR MARVEL, a National Centre of Competence in Research, funded by the Swiss National Science Foundation, Switzerland (grant number 205602) and support from the SNSF through project number 215178. We also acknowledge Yann L. Müller for the implementation of the first version of \texttt{pyeCE} \cite{muller_pyece_2026}.

\section{Author Contributions}
ARN proposed and supervised the project. DKJL implemented the extension of the \texttt{pyeCE} package, performed the analysis, and processed the data. Both authors contributed to the writing and the revision of the manuscript.

\bibliography{references}
\end{document}